# a Monolithic Topologically Protected Phononic Circuit

Si-Yuan Yu[1,2][†], Cheng He[1,2][†], Zhen Wang[1], Fu-Kang Liu[1], Xiao-Chen Sun[1], Zheng Li[1], Ming-Hui Lu[1,2]*, Xiao-Ping Liu[1,2]*, and Yan-Feng Chen[1,2]*

1. *National Laboratory of Solid State Microstructures & Department of Materials Science and Engineering, Nanjing University, Nanjing, Jiangsu 210093, China*
2. *Collaborative Innovation Center of Advanced Microstructures, Nanjing University, Nanjing, Jiangsu 210093, China*

*Correspondence to: luminghui@nju.edu.cn, xpliu@nju.edu.cn and yfchen@nju.edu.cn.*
[†]*These authors contributed equally to this work.*

**Abstract:** Precise control of elastic waves in modes and coherences is of great use in reinforcing nowadays elastic energy harvesting/storage, nondestructive testing, wave-mater interaction, high sensitivity sensing and information processing, etc. All these implementations are expected to have elastic transmission with lower transmission losses and higher degree of freedom in transmission path. Inspired by topological states of quantum matters, especially quantum spin Hall effects (QSHEs) providing passive solutions of unique disorder-immune surface states protected by underlying nontrivial topological invariants of the bulk, thus solving severe performance trade-offs in experimentally realizable topologically ordered states. Here, we demonstrate experimentally the first elastic analogue of QSHE, by a concise phononic crystal plate with only perforated holes. Strong elastic spin-orbit coupling is realized accompanied by the first topologically-protected phononic circuits with both robustness and negligible propagation loss overcoming many circuit- and system-level performance limits induced by scattering. This elegant approach in a monolithic substrate opens up the possibility of realizing topological materials for phonons in both static and time-dependent regimes, can be immediately applied to multifarious chip-scale devices with both topological protection and massive integration, such as on-chip elastic wave-guiding, elastic splitter, elastic resonator with high quality factor, and even (pseudo-)spin filter.

*Introduction*: Topology, the concept originated from modern mathematics, has been playing a very important role in condensed mater physics since the quantum Hall effect was firstly observed in 1980s. In past decades, a set of topological states have been discovered in various electronic systems including quantum Hall effect (QHE), quantum spin Hall effect (QSHE) and topological insulators (TIs)[1-8], whose typical topological feature is characterized by the insulating bulk energy bands but the conducting edge bands, leading to unidirectional charge or spin-dependent propagation at its boundary. Following the generic topological concept and band theory of artificially periodical materials, the topological design and implementation can be extended from fermi to boson system, inspiring lots of bosonic topological states for classical waves, such as topological photonics[9-21], acoustics[22-29] and mechanics[30-33]. The essential impetus is the backscattering-immune edge states, for transportation of bosonic photons/phonons does not need any potential gradients or spin pumps like fermi electrons. Thereby it offers a reflection-free transportation with unparalleled tolerance towards any "non-magnetic" defects and fabrication imperfections, opening up a new era of large-scale photonic/phononic circuits.

Without considering "spin", photonic and fluid acoustic counterparts of QHE with one-way single polarized electromagnetic wave and longitudinal sound wave propagation can be directly analogized. The typical ways to break time-reversal (T) symmetry are using magneto-optical materials in the presence of external magnetic field for photons or moving background for fluid sound. These systems are hard to scalable, hindering their further applications. However, to obtain T-invariant photonic and acoustic QSHE or TI, is not that straightforward. This is because the T operators of bosons (squares to +1) and fermions (squares to -1) are totally different. The intrinsic T symmetry of photons and phonons cannot naturally guarantee Kramers degeneracy and protect topological states. The key factor to realize bosonic QSHE and TI is utilizing artificial symmetry to construct fermi-type pseudo T symmetry (squares to -1). Therefore, the native spins of photons and phonons (with spin-1 or spin-0) act as pseudo-spins in these systems, which will have some kind of property of spin-½ particles.

Elastic systems possess key advantages in RF signal processing[34] enabled by much smaller functional device footprint and strong phonon–phonon interaction. Unfortunately, these systems have slow group velocities, high densities of states, and very large acoustic impedance mismatch. These all make them vulnerable to defects/disorders and thus promote backscattering.

However, though the elastic topological phases have so far been predicted, none has been practical realized. Even in whole acoustics fields, only few phononic topological phases have been actually realized until just recently, e.g., He et al. utilizing accidental degenerated double Dirac cone[28] and Peng et al. using ring-coupling sound circulation[29], both in a scalar (longitudinal wave) system of sound in air. For the elastic wave in solid-state, Mousavi et al. theoretically predicted a double Dirac-cone achieved by symmetric and anti-symmetric Lamb modes using a deep sub-wavelength meta-structures with meticulously engineering[27]. Hence, it still remains a challenge to realize phononic topological phases for a general monolithic solid structure that supports all three elastic polarizations and is scalable to operate at GHz and beyond. Here, we show, other than using former predicted deep sub-wavelength structure[27], which may be herculean task for practical realization, an accidentally double Dirac cone (four-fold degenerated acoustic states) can also be constructed in a plane solid plate, i.e. a plate phononic crystal consisting of identical perforated holes, by simply modulate the geometric correlation of the plate thickness and radii of the holes. These four states can be further hybridized to form two elastic spin-½ states, then to realize elastic QSHE an TI. Our elastic TI with topological protection against a variety of structural imperfections and disorders has opened door for promising applications. Moreover, the physical model demonstrated here on a monolithic solid platform is fully scalable to high frequency applications, leading to great possibility in realizing chip-scale electro-acoustic/opto-acoustic devices with unprecedented backscattering immune transport of SAW, e.g., on a COMS compatible phononic platform.

*Elastic band inversion in a plate phononic crystal*: In bosonic systems, artificial spin-½ states (herein referred as spin states) can be emulated through polarization or modal hybridization. For instance, two degenerate modes M1 and M2 can be hybridized to construct two degenerate spins: spin+/− ≡**M1**+i**M2**/**M1**−i**M2**. These two spins are bosonic by nature and thus still satisfy M1+iM2↔M1−iM2 under time-reversal T=$T_b$ symmetry ($T_b^2$=+1). To construct pseudo T=$T_p$ symmetry ($T_p^2$=−1), under which **M1**→**M2** while **M2**→−**M1**, an additional artificial symmetry is required (*25, 26*). This can be realized by adding magneto-electric coupling with degenerated electric (**E**) and magnetic (**H**) components such that **E**→**H**→−**E**. Here, four-fold degenerate states, instead of a pair of degenerate states, are used to construct bosonic spin states and associated TI. Thanks to $C_{6v}$ symmetry, these required states can be formed accidentally at a double Dirac point via geometry tuning in a unit cell. As shown in Fig.1, the tuning procedure starts with a hexagonal non-bravais lattice containing six identical "atoms" and then gradually changes the distance *b*, measured from the center of the perforated holes to the center of hexagonal unit-cell. When *b* is relatively small, for example *b*=$a_0$ (in this case, a standard honeycomb lattice), there exists band-gap for Lamb-waves, *i.e.*, the SAWs of our interests. Note that though there are stills modes in the band-gap, these modes can be safely ignored for their shear-horizontal (SH) nature. Two pairs of two-fold degenerate surface acoustic states exist at the Brillouin zone center, corresponding to $p_x$/$p_y$ modes and $d_{x2−y2}$/$d_{xy}$ modes, which are similar to *p* and *d* orbitals of electrons. Here $p_x$ obeys symmetry $\sigma_x$/$\sigma_y$=−1/+1; $p_y$ obeys +1/−1; $d_{x2−y2}$ obeys

+1/+1; and $d_{xy}$ obeys −1/−1. Here, $\sigma_{x(y)}$ =+1, −1 represents the even or odd symmetry along the *x* or *y* axis, respectively. When *b* is increased to $1.0873a_0$ in our case, the band-gap vanishes, resulting in an accidental double Dirac point with the desired four-fold degeneracy. Keeping increasing *b*, *e.g.*, to $b=1.12a_0$, results in energy band inversion, characterized by the flipping of the *p* and *d* modes. Consequently, the surface phononic crystal experiences a topological transition from a topological trivial crystal (TC) to a topological non-trivial crystal (nTC).

*Elastic helical edge states*: The topological properties of our system can be exemplified by an interface where topological transition between an nTC and a TC occurs as shown in Fig. 2a. The projected dispersion is shown in Fig. 2b. Ignoring the dispersion of SH modes denoted by grey lines, a single Dirac cone, formed by two gapless dispersions, appears at $k_∥=0$ inside the overlapped bulk gap of both crystals. These two gapless dispersions represent helical edge states, originated from the elastic spins of the bulk states of these two topologically different crystals. At the interface, *p* and *d* states in bulk hybridize to form a pair of normal modes, *i.e.*, one symmetric mode $S = (p_x + d_{x2-y2})/\sqrt{2}$ and one anti-symmetric mode $A = (p_y + d_{xy})/\sqrt{2}$ as shown in Fig. 1c (here the symmetry axis is the normal direction of the interface). These two normal modes are used as the basis to construct two elastic spin-½ states, $S+iA$ and $S−iA$, which satisfy a pseudo time-reversal symmetry related with our system $T_p$ ($T_p^2=−1$). This point is confirmed by checking the result of $T_p$ operating on the basis (mode S and A). Specifically, using a series of $T_p$ operation, the basis is transformed continuously according to " +S → +A → −S → ⋯ ". Therefore, the two states constructed from the basis S and A satisfy Kramers degeneracy, similar to the case of electron spins, but under a pseudo time-reversal symmetry $T_p$.

*Spin-selective waveguide coupler*: Generally, it is uneasy to selectively excite a particular spin in experiment even with multiple excitation sources. Here, we design experimentally a spin-selective waveguide coupler, which allows us to distinguish and separate spin-dependent transport with a very high fidelity even with an unknown spin excitation. As shown in Fig. 3a, our waveguide coupler is divided into three sections by a sandwiched structure composed of TC and nTC. Hence, two topological waveguides are formed: a straight waveguide on the left sided by a partial hexagonal waveguide on the right. There are four input/output ports in the whole waveguide coupler, two in the straight waveguide are labeled as P1 and P2, and two in the partial hexagonal waveguide are labeled as P3 and P4, respectively. Coupling of nearby waveguides can generally happen if both of them support the same eigenmodes. In the coupling of two ordinary waveguides, the direction of wave propagating from one waveguide to the other will be chosen principally at the direction of original energy flow. For example, if the energy flows initially from bottom to top along the straight waveguide, when the wave is transporting near the coupling area of the two waveguides, the energy will instinctively flow into the hexagonal waveguide follow its original direction, i.e., flows into the upper half of the hexagon waveguide in a clockwise direction. Conversely, if the original energy is from top to bottom in straight waveguide, the energy will flow into the lower half of the hexagonal waveguide in an anti-clockwise direction after the coupling happened. However, in a spin-dependent waveguide, though both spin+ and spin- mode is supported, each of them can only flows along the unique direction, Put simply, if people place a transducer near one port of the waveguide, only the definite spin+ or spin- mode can be excited and propagate through the waveguide. For example, if a transducer were placed near the bottom (top) port of the straight waveguide P1 (P2), only elastic spin+ (spin−) can be excited and transmit through the waveguide with respect to its transmission direction, i.e., nTC is located on the left (right) side while TC is located on the right (left) side. At this circumstance, because both the elastic spin+ and spin- is preserved in the coupling of two waveguides and no spin inversion mechanism is involved in whole configuration, only the anti-clockwise (clockwise) propagation of elastic wave in spin+ (spin-) mode will appear in the hexagonal waveguide in final, implying that the energy will no longer flows along its original direction after passing from one waveguide to the other.

Elastic energy distributions by both finite element method (FEM) simulation and practical experiment with a longitudinal wave transducer attached to the sample substrate near P1 and P2 are shown in Fig. 3b and Fig. 3c, respectively. At an ultrasonic frequency in the bulk band-gap, it is clear that there are only one elastic spin in the clockwise or counterclockwise direction can be detected in the hexagonal waveguide. Such observations are consistent with our theoretical analysis above. Moreover, spectra of the energy densities near both port P3 and P4 (defined as $E_{P3}$ and $E_{P4}$) are also measured. Their contrast ratio, defined as $(E_{P3}-E_{P4})/(E_{P3}+E_{P4})$ are shown in Fig. 3d and 3e, respectively, confirming a high contrast energy flow within the bulk band-gap frequency where the waveguide coupling happens. All these experimental observations for the spin-selective waveguide coupler model confirmed clearly the unique one-way spin-dependent transportation in our elastic topological insulator.

*Robust, flexible and low-loss elastic wave transportation*: Our elastic topological edge states are symmetry protected with topological immunity against all non-spin-mixing defects. Experimentally, different defects are intentionally introduced into our topological waveguide of the nTC-TC interface to study the robust propagation for elastic wave. The lower panels of Fig. 4a-d show schematics of the samples, in which the geometries of the elastic topologically-protected waveguides are indicated by the yellow dashed lines. Two different defects, including a cavity with random missing perforated holes (Fig. 4b) and a random disordered lattice of perforated holes (Fig. 4c) are implemented in an original straight waveguide without defect (Fig. 4a). Besides, an "Z"-type waveguide with two 120° sharp bends (Fig. 2d) is also made for our test. Note in all the cases, no spin-mixing mechanisms (effective "magnetic" impurities) are presented intentionally to break the topological protection. Elastic energy distributions by both simulations and practical measurements with a spin+ elastic wave incidence corresponding to the above cases are shown in the upper panels of Fig. 4a-d with measured transmission spectra shown in Fig. 4e, verifying the robustness and flexibility of this waveguide, for the elastic wave detouring around all these arbitrary defects and bends while maintains nearly a loss-free elastic transmission. We especially note that the bends robustness is not only for the 120° bends (as in our experiments), but also for any arbitrary bends thanks for our band inversion is achieved in the whole reciprocal space. In comparison, an original waveguide made of lines cavities in the exactly same TC in Fig. 1a (or nTC in Fig. 1c) is also constructed for a control experiment. The results for the similar defects and sharp bends are drastically different. The defects will cause distinct elastic resonances, while the sharp bends will severely inhibits the elastic forward propagation, both leading to a decreased transmission or even a total reflection. Our proposed phononic waveguide clearly distinguishes itself from traditional phononic transmission lines with unparalleled advances in imperfection immune and arbitrary shape, especially it is for the elastic wave in a monolithic electro-acoustic substrate with simple implementation and extremely low intrinsic loss, which is highly expected but has never been achieved even closely before.

*Summary*: We have confirmed the elastic spin-½ states and the elastic QSHE in a plate phononic crystal. Our configuration provides a general and convenient implement for elastic-wave system, which supports three polarizations. Thanks to its easy-fabricate geometry, static nature, the external "force"-free setup and non-exotic materials, such monolithic solid structure can be scalable to operate at GHz and beyond. We believe that our demonstration of plate elastic topological phenomena here can immediately offer promising applications such as on-chip elastic wave-guiding, splitter, resonator with high quality factor, and even (pseudo-)spin filter.

**Acknowledgments**

The work was jointly supported by the National Key R&D Program of China (2017YFA0303700 and 2017YFA0305100), National Basic Research Program of China (2013CB632904, 2013CB632702, 2015CB659400), and the National Nature Science Foundation of China (Grant No. 11625418, No. 11474158, No. 51472114, and No. 61378009). We also acknowledge the support of the Natural Science Foundation of Jiangsu Province (BK20140019 and BK20150057) and the support from the Academic Program Development of Jiangsu Higher Education (PAPD).

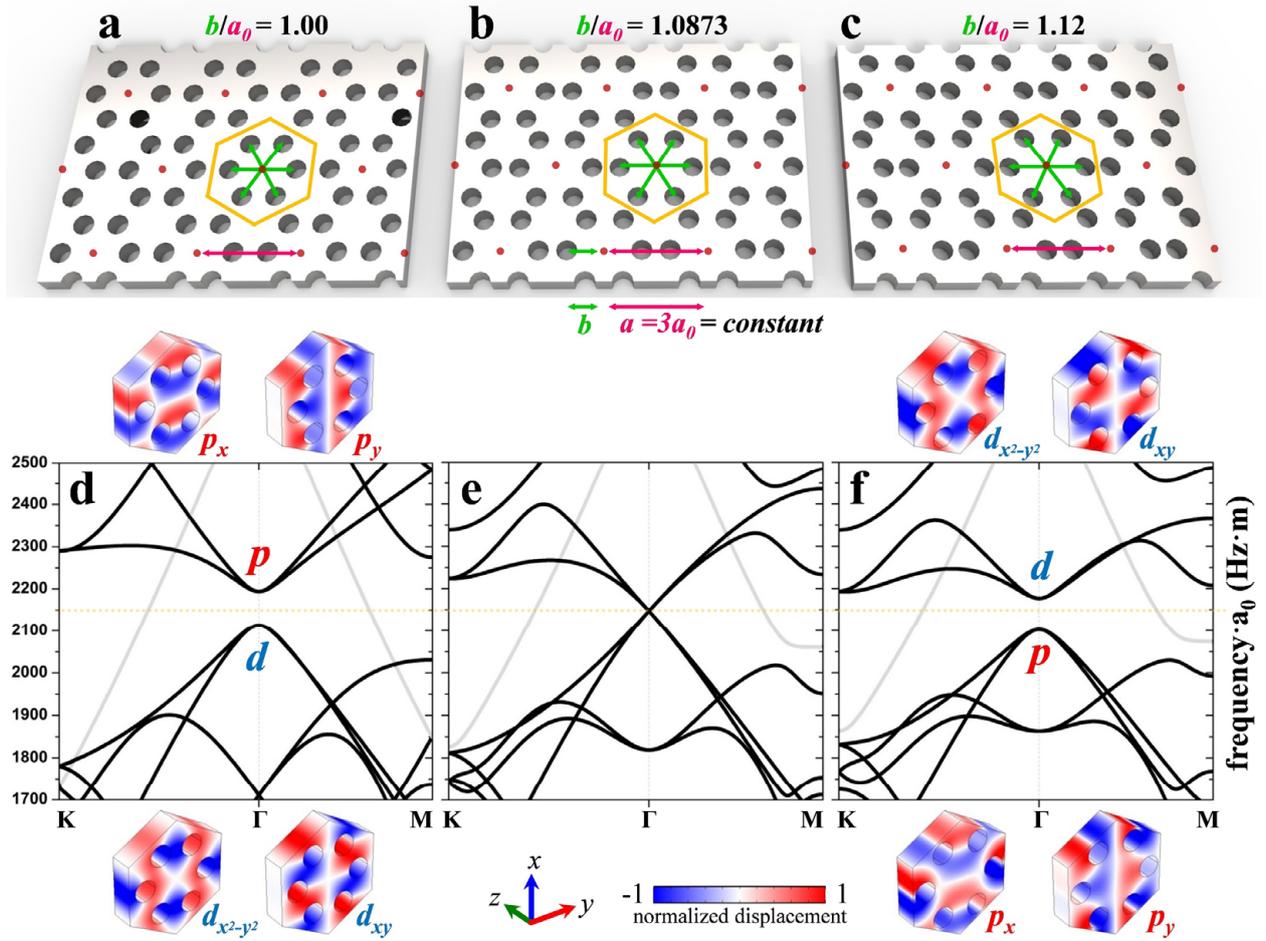

**Fig.1. Elastic topological band inversion via geometrical revolution.**

**a**, **b** and **c**: geometrical revolution of a plate of elastic phononic crystal composed of identical perforated holes in triangular symmetry. In the revolution process, thickness of the substrates equals to $0.4a_0$, lattice constant a (equals to $3a_0$) and radius of perforated holes r (equals to $0.18a_0$) maintain constant. However, inside each unit-cell of the crystal, as the yellow hexagons indicate, the length between the center of the perforated holes to the center of the unit-cell b increases form $1.00a_0$ (**a**), $1.0873a_0$ (**b**) to $1.12a_0$ (**c**), respectively.

**d**, **e** and **f**: full band diagrams corresponding to the above three cases, respectively, illustrating a band inversion process. In the original lattice (**a**) $b=1.000a_0$, there presnets two two-fold degenerated states, which can be denoted by $p_x/p_y$ and $d_{x2-y2}/d_{xy}$ and are separated by a band-gap. Note although there are still modes in the band-gap, as high-lighted grey modes/dispersions in these diagrams, these modes/dispersions are essentially independent/irrelevant form others (both in vibration features and topological properties), and could be selectivity inhibited by some specified manners. When b increases to a unique value, like (**b**) $1.0873a_0$, the band-gap disappears and an accidental double Dirac cone with a four-fold degeneracy node is formed. When b further increases, like to (**c**) $1.12a_0$, the band-gap reappears along with two inverted two-fold degenerate elastic states, corresponding to $d_{x2-y2}/d_{xy}$ and $p_x/p_y$. In the whole geometrical revolutions, an elastic topological phase transition occurs at the four-fold degeneracy node by examining Berry phases of the bands which forming the band-gaps. Insets which surround the band diagrams are displacement distributions of the degenerated elastic eigenstates in (out-of-plane) z directions.

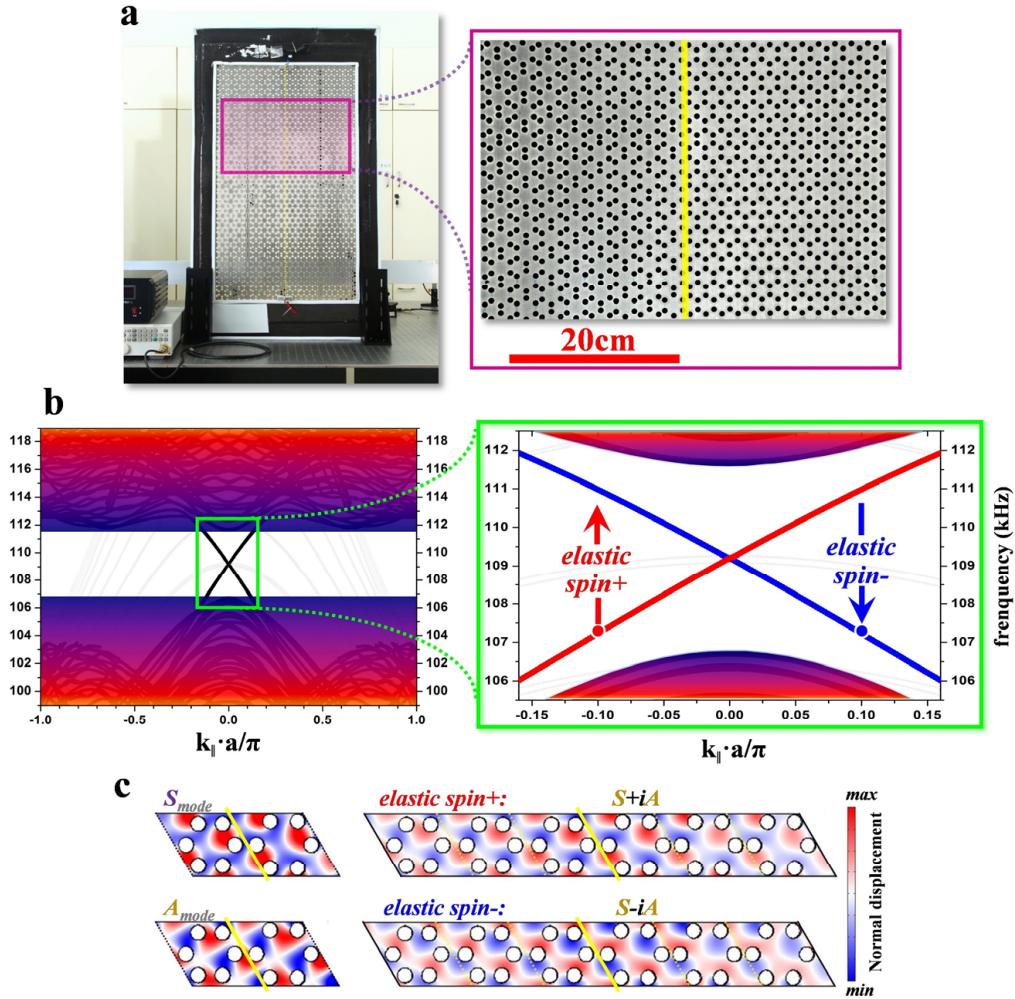

**Fig. 2. Elastic helical edge states.**

**a:** Sample of elastic topologically-protected edge/waveguide constructed by stacking a topological trivial crystal (Fig.1a, b=1.00$a_0$) with a topological non-trivial crystal (Fig.1c, b=1.12$a_0$).

**b**: Calculated projected elastic band diagram of the topologically-protected phononic circuit constructed by stacking a topological trivial crystal (**Fig.1a**, $b=1.00a_0$) with a topological non-trivial crystal (**Fig.1c**, $b=1.12a_0$). Zoomed-in band diagram illustrates two helical edge states representing two spin edge-states: elastic spin+ and elastic spin-. Grey dispersions here are the shear-horizontal modes.

**c**: Representative examples of the field distributions of the elastic displacement vertical to the substrate for both S mode, A mode, and their hybridizations.

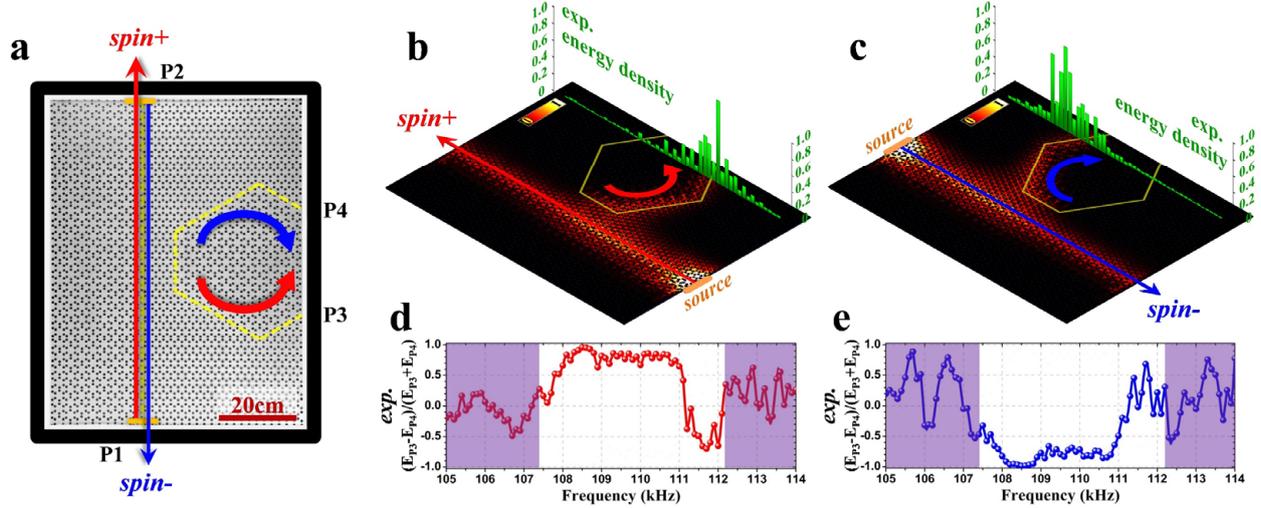

**Figure 4 | Elastic one-way spin-dependent transport.**

**a**: Photograph of an elastic waveguide-coupler to demonstrate elastic one-way spin-dependent transport. The whole sample is divided into three sections by a sandwiched-like structure composed of a topological trivial phononic crystal (**Fig.1a**, $b=1.00a_0$) between two topological non-trivial phononic crystals (**Fig.1c**, $b=1.12a_0$). Thus, two topological protected edges/waveguides are formed: a straight edge/waveguide on the left, sided by a partial hexagonal edge/waveguide on the right, as the yellow dotted lines indicate. Two ports of the straight edge/waveguide are labeled as P1 and P2, and two ports of the partial hexagonal edge/waveguide are labeled as P3 and P4, respectively. In this experiment, when the coupling of the two waveguides happens, only anti-clockwise (clockwise) circulating propagation along the partial hexagonal edge/waveguide is allowed for elastic spin+ (spin−) as indicated by the red (blue) circular arrow.

**b** and **c**: (bottom) Heat maps are calculated elastic energy-density fields at a frequency (108.5kHz) within the bulk band-gap, (top) Green bars are experimental results spatially measured along a straight line near both P3 and P4 of the partial hexagonal edge/waveguide. In both the calculation and experiment, we applied ultrasonic longitudinal-wave transducers attached to the sample surface at P1 and P2 to excite only the elastic spin-up and spin-down edge mode, respectively.

**d** and **e**: Experimental measured spectra of contrast ratio of elastic energy-density between P3 and P4, defined as $(E_{P3}-E_{P4})/(E_{P3}+E_{P4})$, for elastic spin+ incidence from P1 and for elastic spin− incidence from P2, respectively. Shadow regions correspond to the bulk bands.

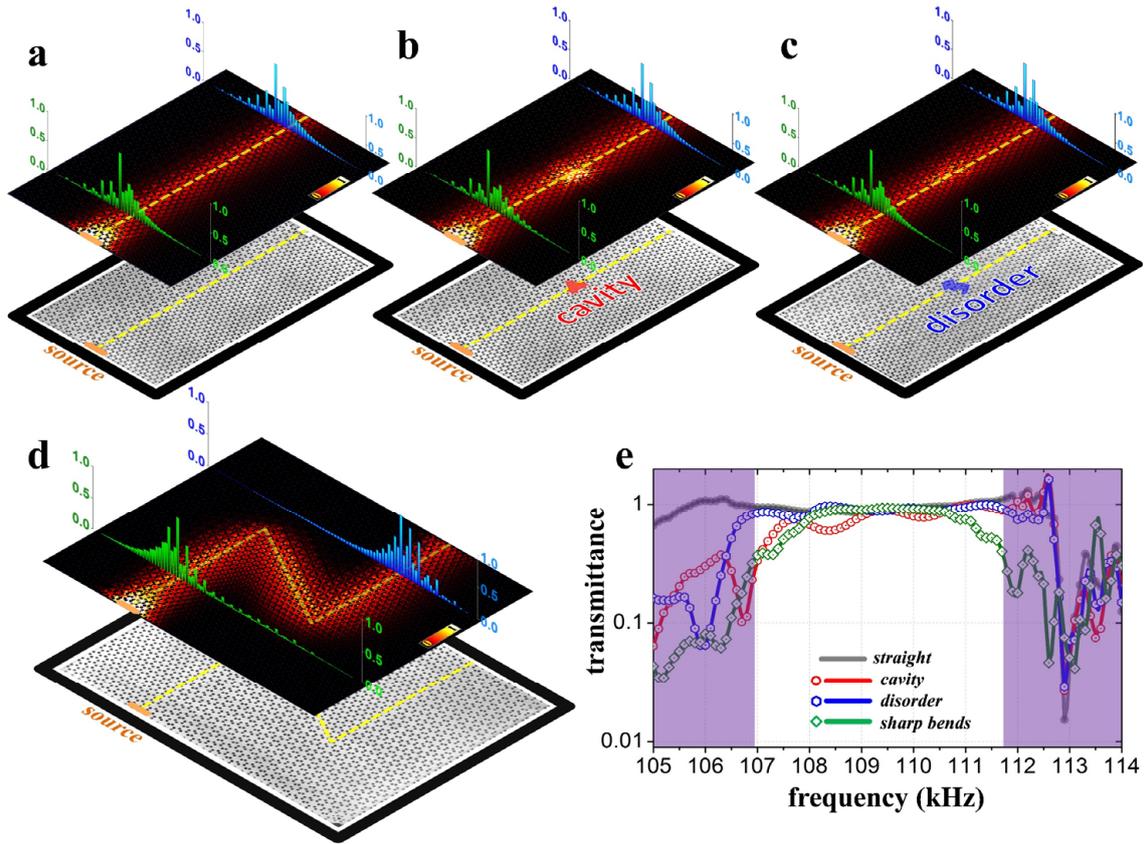

**Fig. 4. Robust and flexible elastic wave transport.**

**a**, **b**, **c** and **d**: (bottom) photographs of different topological-protected edges/waveguides to demonstrate robust and flexible elastic wave transport. The yellow dotted lines indicate locations of the edges/waveguides constructed by stacking a topological trivial phononic crystal (**Fig.1a**, $b=1.00a_0$) with a topological non-trivial phononic crystal (**Fig.1c**, $b=1.12a_0$). Edge/waveguides with (**a**) no defect and three different defects of (**b**) an arbitrary-shaped cavity without perforated holes, (**c**) an arbitrary-shaped disorder of random ordered perforated holes (**d**) a 'Z' type bends are fabricated. (middle) Heat maps are calculated elastic energy-density fields for elastic spin-up incident at a frequency (110.0kHz) within the bulk band-gap, corresponding to the four different configurations, respectively. (top) Green and blue bars are experimental results spatially measured along two straight lines cut through the edges/waveguides before and behind the defects, respectively.

**e**: Experimental measured transmission spectra for these elastic topological protected edges/waveguides. The black curve corresponds to the case without any defects as in **a**, while the red, blue and green curves correspond to the case of **b**, **c** and **d**, respectively. Shadow regions correspond to the bulk bands.